\documentstyle[pre,aps,multicol,epsf]{revtex}
\renewcommand{\narrowtext}{\begin{multicols}{2} \global\columnwidth20.5pc}
\renewcommand{\widetext}{\end{multicols} \global\columnwidth42.5pc}

\newcommand{\Rrule}{\vspace{-0.1in}\hfill\vrule depth1em height0pt \vrule
  width3.5in height.2pt depth.2pt\vspace*{-0.1in}}
\def\bml{\begin{mathletters}}
\def\eml{\end{mathletters}}
\def\beq{\begin{equation}}
\def\eeq{\end{equation}}
\def\bea{\begin{eqnarray}}
\def\eea{\end{eqnarray}}
\def\pa{\partial}
\def\e{{\rm e}}
\def\tr{{\rm tr}}

\def\sgn{{\rm sgn}}
\def\l{\lambda}
\def\q{\tilde{q}}
\def\p{\tilde{p}}
\def\R{\tilde{R}}
\def\r{{\bbox{\rm r}}}
\def\ii{{{\bbox{\rm i}}}}
\def\i{{\hat{\bbox{\rm i}}}}
\def\x{{\hat{\bbox{\rm x}}}}
\def\y{{\hat{\bbox{\rm y}}}}
\def\z{{\hat{\bbox{\rm z}}}}
\def\s{\sigma}

\begin{document}
\draft
\title{ 
Level spacings at the metal-insulator transition
in the Anderson Hamiltonians
and multifractal random matrix ensembles
}
\author{Shinsuke M. Nishigaki${}^*$}
\address{
Institute for Theoretical Physics, University of California at Santa Barbara,
Santa Barbara, California 93106
}
\date{September 9, 1998}
\maketitle
\begin{abstract} 
We consider orthogonal, unitary, and symplectic
ensembles of random matrices with $(1/a)(\ln x)^2$ potentials,
which obey spectral statistics different from the 
Wigner-Dyson and are argued to have multifractal eigenstates.
If the coefficient $a$ is small,
spectral correlations in the bulk are universally governed by a 
translationally invariant, one-parameter generalization of the sine kernel.
We provide analytic expressions
for the level spacing distribution functions of this kernel,
which are hybrids of the Wigner-Dyson and Poisson distributions.
By tuning the single parameter,
our results can be excellently fitted to the numerical data
for three symmetry classes of the three-dimensional Anderson Hamiltonians
at the metal-insulator transition, previously measured
by several groups using exact diagonalization.
\end{abstract}
\narrowtext

\section{introduction}
Quantum mechanics described by stochastic ensembles of Hamiltonians \cite{Por},
and by Hamiltonians with classically chaotic trajectories \cite{BGS}, 
have been a subject of intense study for years.
In contrast to Hamiltonians of classically integrable systems
whose energy levels are mutually uncorrelated,
chaotic Hamiltonians generally exhibit strong correlation among
levels.
To simulate this level repulsion in chaotic or disordered Hamiltonians, 
Wigner introduced the random matrix ensembles (RME) \cite{Meh}.
In the RME defined as an integral over $N\times N$ matrices,
only the antiunitary symmetry of the Hamiltonian is respected,
and its spatial structure is completely discarded.
Despite this extreme idealization, 
analytic predictions from RME's beautifully explain
the spectral statistics of chaotic or disordered Hamiltonians \cite{GMW}.
This success is accounted for by the fact that the RME is not merely 
an idealization of disordered Hamiltonians but 
it is indeed equivalent to the latter
under a situation where the mean energy level spacing $\Delta$ 
is much smaller than
the Thouless energy $E_c$ (inverse classical diffusion time)
that the dimensionality does become unimportant
\cite{Efe}.
This on the other hand implies that in a region where
the mean level spacing is equal to or larger than the Thouless energy,
and the associated states tends to localize due to diffusion,
standard RME's cannot provide good quantitative descriptions.

As a concrete example of systems of disordered conductors,
let us take the Anderson tight-binding Hamiltonian (AH)
\cite{And,KMK},
\beq
H=\sum_\r \varepsilon_\r  a_\r^\dagger a_\r
+\sum_{\langle \r,\r' \rangle} a_\r^\dagger a_{\r'},
\label{AH1}
\eeq
describing free electrons in a random potential.
Here $a_\r^\dagger$ and $a_\r$ are creation and annihilation operators
of an electron at a site $\r$
on a three-dimensional (3D) toroidal lattice of size $L^3$, 
$\varepsilon_\r$'s are site energies
that are mutually independent stochastic variables
derived from a uniform
distribution on an interval $[-W/2, W/2]$, and 
the sum $\langle \r,\r' \rangle$ 
is over all pairs of nearest-neighboring sites.
The antiunitary symmetry of the AH [Eq.\ (\ref{AH1})] is orthogonal ($\beta=1$),
as it respects the time-reversal symmetry and has no spin dependence.
For small values of disorder $W$, all eigenstates are extended.
An overlap of such extended states 
$\Psi_i(\r)$ and $\Psi_j(\r)$
is expected to induce repulsion between 
associated eigenvalues,
of the form $|x_i-x_j|$. 
Then the statistical fluctuation of
these eigenvalues is described well by the Gaussian orthogonal ensemble
of random matrices 
whose joint probability distribution consists of
the product of a Vandermonde determinant $\prod_{i<j}|x_i-x_j|$
and Gaussian factors $\prod_{i}\e^{-x_i^2}$.
The quadratic potential in the latter is merely for the sake
of technical simplicity, and the deformation of the potential by 
generic polynomials universally leads 
to the same Wigner-Dyson statistics \cite{BZ}.
As we increase the disorder $W$, 
states associated with eigenvalues close to the edges of the spectral band
are believed to start localizing.
When the disorder is as large as to induce 
the metal-insulator transition (MIT) $E_c/\Delta\simeq 1$,
eigenstates are observed to be multifractal \cite{SG},
characterized by 
an anomalous scaling behavior of the moments of 
inverse participation ratio \cite{Weg,CP,Jan}
\beq 
\sum_\r \langle 
|\Psi_{i}(\r)|^{2p} \rangle
\propto L^{-D_p (p-1)}. 
\label{mf}
\eeq
This property implies a slowly decreasing overlap between states \cite{Cha}
(for $|x_{i}-x_{j}|\gg \Delta$),
\beq
\sum_\r \langle 
|\Psi_{i}(\r)|^2 |\Psi_{j}(\r)|^2 \rangle
\propto |x_{i}-x_{j}|^{-(1-{D_2}/{d})} .
\label{multi}
\eeq
For large enough disorder, the off-diagonal (hopping) term
of the AH becomes negligible,
leading to mutually uncorrelated eigenvalues.
Each eigenstate is almost localized to a single site. 
Thus the spectrum of the AH shows a gradual crossover from the 
Wigner-Dyson to the Poisson as one increases the disorder $W$,
while keeping the size $L$ fixed finite.
A characteristic observable in these studies of spectral statistics is
the probability $E(s)$ of having no eigenvalue
in an interval of width $s$, or equivalently
the probability distribution of 
spacings of two consecutive eigenvalues
$P(s)=E''(s)$.
These observables 
capture the behavior of local correlations of 
a large number of energy eigenvalues,
as the former consists of an infinite sum of integrals of
regulated spectral correlators
(the subscript reg denotes its regular part, i.e.,
with $\delta$-functional peaks at coincident $x_i$'s subtracted),
\bea
&&E(s)\!=\!\sum_{n=0}^\infty \frac{(-1)^n}{n!}\!
\int_{-s/2}^{s/2}\!\! dx_1\cdots dx_n
\left\langle \rho(x_1)\cdots \rho(x_n) \right\rangle_{\rm reg} ,
\!\!\!
\label{Espert}
\eea
and are more conveniently measured by 
the exact diagonalization of random Hamiltonians than the spectral 
correlators. 
The Poisson distribution is characterized by 
$P_{{\rm P}}(s)=\exp(-s)$, and the Wigner-Dyson distribution
is well approximated by the Wigner surmise
$P_{{\rm W}}(s)=(\pi s/2)\exp(-\pi s^2/4)$.

Recent technical developments 
\cite{SSSLS,Eva,EK,HS,HS94,VHSP,Hof96,Hof98,ZK95,ZK97,BSZK,KSOS,BM}
on the exact diagonalization of the AH on a large size of lattices 
prompted
analytical studies on its spectrum \cite{AZKS,KLAA,AKL,AM,FM}.
It was noticed in Ref.\ \cite{SSSLS}
that at the MIT point
with disorder $W\sim 16.5$,
the level spacing distribution function (LSDF) $P(s)$
is independent of the size $L$.
From this finding these authors have argued that
in the thermodynamic limit
there exist only three universality classes:
Wigner-Dyson, Poisson, and the third, critical statistics.
The presence of critical LSDF's was also observed
for the unitary ($\beta=2$) and symplectic ($\beta=4$) cousins,
i.e., AH's under a magnetic field \cite{HS94,Hof96,BSZK},
\bea
&&H=\sum_\r \varepsilon_\r  a_\r^\dagger a_\r
+\sum_{\langle \r,\r' \rangle} V_{\r\r'}a_\r^\dagger a_{\r'},
\label{AH2}\\
&& 
V_{\r,\r\pm\x}=\e^{\mp2\pi i \alpha {\rm r}_y},\ 
V_{\r,\r\pm\y}=V_{\r,\r\pm\z}=1, \nonumber
\eea
and 
with spin-orbit coupling \cite{EK,KSOS,Hof98},
\bea
&&H=\sum_{\r,\s=\pm} \varepsilon_{\r}  a_{\r\s}^\dagger a_{\r\s}
+\sum_{\langle \r,\r' \rangle, \s,\s'} 
V_{\r\s,\r'\s'}a_{\r\s}^\dagger a_{\r'\s'},
\label{AH4}\\
&& V_{\r\s,\r\pm\i\,\s'}=
\left(\e^{\mp i\theta \bbox{\sigma}_{\ii}}\right)_{\s\s'}
\ \ \ \ (\i=\x, \y, \z).\nonumber
\eea
The critical LSDF's are found to be independent of
the strength of 
the magnetic field $\alpha$ or the spin-orbit coupling $\theta$.
For all values of $\beta$, the Wigner-Dyson-like behaviors
$P_{{\rm (cr)}}(s) \propto s^\beta$ for small $s$
have been confirmed.
There were disputes over the large $s$ asymptotic behavior
of the LSDF's,
but accurate measurements on large lattices \cite{ZK97} 
strongly support 
the Poisson-like behavior 
$P_{{\rm (cr)}}(s) \sim \exp(-{\rm const}\times s)$ for large $s$,
excluding a nontrivial exponent
$P_{{\rm (cr)}}(s) \sim \exp(-{\rm const}\times  s^{1+\gamma})$
predicted in Ref.\ \cite{AKL}.
In the light of the success of the random matrix (RM) description of
extended states, a natural resort to describe this critical statistics
is to consider a deformed RME
that violates the above mentioned universality of the 
Wigner-Dyson statistics
for the Gaussian ensembles.
The validity of the RM description for such critical systems
is of course far from clear,
because the existence 
of the MIT crucially relies upon the dimensionality, whereas the RME 
has no spatial structure.
Nevertheless,
under an assumption that the spectra of the AH be described 
by RME's, an attempt \cite{APM} was made to reconstruct a
random matrix potential out of the 
macroscopic spectra of the AH. 
There, it was observed that a potential of the form 
\beq
V(x)=\frac{1}{2a} [\ln (1+b |x|)]^2
\label{VlogAB}
\eeq
explains well the numerical data.
The above potential indeed violates the universality of 
the Wigner-Dyson statistics
that is guaranteed only for polynomial potentials \cite{CWK}.
This observation leads to a speculation that 
the critical level spacing distribution might be
derived from a RME with a potential (\ref{VlogAB}).
Later the LSDF of orthogonal RME's of type 
(\ref{VlogAB})
has been measured by using the Monte Carlo simulation \cite{Can} 
in order to compare it with that from exact diagonalization of
the AH \cite{ZK95}.
Excellent agreement between the two was found
by tuning the parameter $a$ to $2.5$.
Motivated by this success, we shall 
derive analytic forms of the LSDF's of
RME's with orthogonal, unitary and symplectic symmetries.
(The unitary case has already been reported 
\cite{Nis}.)
In doing so, we shall retain all perturbative (polynomial in $a$)
parts of the spectral kernel,
and discard unphysical nontranslationally invariant parts
of order $O(\e^{-{\pi^2}/{a}})$.

This paper is organized as follows.
In Sec.\ II we review 
nonstandard features of RME's with log-squared potentials.
In Sec.\ III we follow the method of Tracy and Widom \cite{TW94} to
derive the LSDF's from an approximated translationally invariant kernel.
In Sec.\ IV we shall compare our results 
with the numerical data of the AH's
with orthogonal, unitary, and symplectic symmetries.
In Appendix A we collect standard results on Fredholm determinants in 
random matrix theories
that are relevant for our purpose.

\section{RME with log-squared potential}
In this section we review properties of
RME's with log-squared potentials.
For small enough $a$, we derive a translationally invariant kernel,
whose level spacing distribution will be our subject in this paper.

We consider $N\times N$ random
real symmetric ($\beta=1$), 
complex hermitian ($\beta=2$), and
quaternion selfdual ($\beta=4$) matrix ensembles,
whose joint probability densities of eigenvalues are given by
\beq
{\cal P}_\beta(\l_1,\ldots,\l_N)
\propto
\prod_{i=1}^N \e^{-V(\l_i)}
\prod_{1\leq i < j \leq N} \left| \l_i-\l_j \right|^\beta .
\eeq
with a potential growing as
\beq
V(\l)\sim \frac{1}{2a} (\ln \l)^2 \ \ 
(\l \gg 1).
\label{Vlog2}
\eeq
The potential is assumed to be regularized at the origin,
as in Eq.\ (\ref{VlogAB}).
For a particular form of the potential
\beq
V(\l)=\sum_{n=1}^\infty \ln[1+2 q^{n} \cosh (2\,{\rm arcsinh}\l)+q^{2n}],
\label{Vq}
\eeq
$(0<q<1)$,
which behaves as Eq.\ (\ref{Vlog2}) with $a=\ln(1/q)/2$,
corresponding orthogonal polynomials are known as
the $q$-Hermite polynomials \cite{IM}.
Using their asymptotic form,
Muttalib {\it et al.} \cite{MCIN}
have obtained the exact kernel 
for the unitary ensemble
with the potential (\ref{Vq})
in the large $N$ limit,
\bea
&&K^{\rm (exact)}(x,y)={\rm const}\times
\frac{\sqrt{\cosh 2a x \cosh 2a y}}{\cosh a(x+y)}\times
\label{KqGUE}\\
&&
\frac{\vartheta_4(x+y,\e^{-{\pi^2 / a}})}{
\sqrt{
\vartheta_4(2x,\e^{-{\pi^2 / a}})
\vartheta_4(2y,\e^{-{\pi^2 / a}})
}}
\frac{\vartheta_1(x-y,\e^{-{\pi^2 / a}})}{\sinh a(x-y)} .
\nonumber
\eea
Here $\vartheta_\nu$
are the elliptic theta functions, and the variable $x\equiv \l/(2a)$
is so rescaled that the average level spacing is unity.
In order to eliminate nonuniversal effects 
involved by the regularization of the potential
in the vicinity of the origin from the bulk correlation,
we need to take
\beq
x, y \gg  1,\ \ \mbox{with}\ \ |x-y|=\mbox{bounded} .
\eeq
Then we obtain an asymptotic form of the kernel
\bea
&&K^{\rm (asympt)}(x,y)\!=\!{\rm const}\times\nonumber\\
&&\frac{\vartheta_4(x+y,\e^{-{\pi^2 / a}})}{
\sqrt{
\vartheta_4(2x,\e^{-{\pi^2 / a}})
\vartheta_4(2y,\e^{-{\pi^2 / a}})
}}
\frac{\vartheta_1(x\!-\!y,\e^{-{\pi^2 / a}})}{\sinh a(x-y)}.
\label{Ktheta}
\eea
The above kernel is still not translationally invariant
due to the reason explained below.
Now we make a further simplification of the kernel
by using an approximation.
For $\e^{-{\pi^2}/{a}}\ll 1$,
we can discard subleading orders from the 
$q$ expansion of the theta functions
in terms of trigonometric functions.
Then we obtain a translationally invariant kernel,
up to $O(\e^{-{\pi^2}/{a}})$,
\beq
K(x,y)=\frac{a}{\pi}
\frac{\sin \pi (x-y)}{\sinh a(x-y)}.
\label{KMCIN}
\eeq
The universality of this deformed kernel
within RME's is observed for
the $q$-Laguerre unitary ensemble \cite{BCM},
the finite-temperature Fermi gas model \cite{MNS},
and subsequently for unitary ensembles whose potentials
have the asymptotics (\ref{Vlog2}) \cite{Can}.
This universality 
can be considered as an extension of the 
universality of the sine kernel \cite{BZ},
\beq
K(x,y)=\frac{\sin \pi (x-y)}{\pi(x-y)} ,
\label{Ksin}
\eeq
for polynomially increasing potentials,
proven via the asymptotic WKB
form of the wave functions
\bea
&&\psi_N(\l) \sim \cos\left(
\pi \int^\l \rho(\l)d\l + \frac{N\pi}{2} \right) , \\
\label{WKB}
&&K(\l,\l')\sim
\frac{\sin[ \pi ( \int^\l \rho - \int^{\l'} \rho)]}{\l-\l'}.
\label{K}
\eea
Here $\rho(\l)$ stands for the exact unnormalized
spectral density, $K(\l, \l)$.
In the spectral bulk of the RME with polynomially increasing potentials, 
the spectral density divided by $N$ is bounded, and is 
locally approximated by a constant
when measuring $\l$ in unit of the mean level spacing.
This slowly-varying function is called the mean-field spectral density
$\overline{\rho}(\l)$, given by \cite{Mus}
\beq
\overline{\rho}(\l) \equiv
\frac{N}{\pi\sqrt{R^2-\l^2}}
+\frac{1}{\pi^2}
-\!\!\!\!\!\!\int_{-R}^R \frac{d \mu}{\l-\mu} \sqrt{{R^2-\l^2}\over{R^2-\mu^2}}
\frac{V'(\mu)}{2} .
\label{mfrho}
\eeq
Here $\pm R$ are the end points of the spectrum,
determined by $\int_{-R}^R \overline{\rho}(\l) d\l=N$.
After replacing the exact ${\rho}(\l)$
by the mean-field $\overline{\rho}(\l)$, the unfolding map
\beq
\l \mapsto x=\int^\l \overline{\rho}(\l)d\l
\eeq
becomes merely a linear transformation, leading 
universally to the sine kernel.
On the other hand, in our case of
the potential (\ref{Vlog2}), the mean-field spectral density
(\ref{mfrho}) behaves as
\beq
\overline{\rho}(\l) \sim \frac{1}{2a |\l|},
\label{rhox}
\eeq
implying an unusual unfolding map 
\beq
\l \mapsto x=\frac{1}{2a}\sgn(\l)\ln |\l| ,
\label{unfold}
\eeq
while the formula (\ref{K}) stays valid \cite{CL}.
Then the kernel (\ref{K}) universally reduces to Eq.\ (\ref{KMCIN})
after this unfolding.

It is clear from the form of the kernel (\ref{KMCIN}) that
a set of eigenvalues with $|x_i-x_j| \gg 1/a$ obeys the 
Poisson statistics, i.e., is uncorrelated.
On the other hand, 
a set of eigenvalues with $|x_i-x_j| \ll 1/a$ obeys the Wigner-Dyson 
statistics,
because Eq.\ (\ref{KMCIN}) is then approximated by the sine kernel,
up to $O(a^2)$.
To be precise, the kernel (\ref{KMCIN}) signifies
the multifractality of the eigenstates \cite{KM}.
To see this, we first note that 
the property (\ref{multi}) of the fractal states
leads to a compressible gas of eigenvalues, i.e.,
a linear asymptotics of the number variance $\Sigma^2$
within an energy window of width $S$ \cite{CKL}
$[Y_2(x)\equiv 1-\langle \rho(x)\rho(0)\rangle_{\rm reg}]$,
\bea
\Sigma^2(S)
&\equiv& S-2\int_0^S dx (S-x)Y_2(x)
\label{Sigma2}
\\ 
&\sim& 
\frac12 \left( 1-\frac{D_2}{d} \right)S\equiv
\chi S  \ \ \ (S\gg 1) .
\label{delta}
\eea
The RME's with the scalar kernel (\ref{KMCIN}) 
indeed enjoy this asymptotic behavior
with the level compressibility $\chi$ given by
\beq
\chi=
\frac{a}{\pi^2} 
+O(\e^{-{2\pi^2 / a}})
\label{chi}
\eeq
for unitary \cite{BCM,KM} and orthogonal ensembles,
and for the symplectic ensemble 
the exponential correction is replaced by $O(\e^{-\pi^2/a})$.
Moreover, the {\it multi\/}fractality of the RME's with the deformed kernel 
(\ref{KMCIN}) has been concluded \cite{KM}
through the equivalence between 
the finite-temperature Fermi gas model 
(having the universal deformed kernel)
and a Gaussian banded RME
that is proven to have multifractal eigenstates \cite{MFDQS}.
This brings forth the possibility
of describing the spectral statistics of the
AH's at the MIT by RME's with the deformed kernel.

We should emphasize an important fact 
that the RME's with log-squared potentials cannot describe 
disordered systems with large disorder.
While the AH in the $W\rightarrow\infty$ limit
leads to the Poisson statistics,
the RME with Eq. (\ref{Vlog2}) does not
obey the Poisson statistics in the limit $a\rightarrow\infty$
\cite{Can,BBP}.
It is because
the joint probability distribution of RME's
after the unfolding (\ref{unfold}) leads to
\bea
&&{\cal P}_\beta(x_1,\cdots,x_N)={\rm const}\times \nonumber\\
&&\prod_{1\leq i < j \leq N} \left|\pm\e^{2a |x_i|} \mp \e^{2a |x_j|}
 \right|^\beta 
\prod_{i=1}^N \e^{-2a x_i^2 } \e^{2a |x_i|}.
\eea
In the limit $a\rightarrow \infty$, 
each factor of $\left|\e^{2a |x_i|} \mp \e^{2a |x_j|} \right|^\beta$
is dominated by an exponential with a larger modulus.
Thus the Vandermonde determinant is approximated by
\beq
\prod_{i=1}^{N} \e^{2a\beta(i-1)|x_i|} \ \ \ \ 
(\mbox{for   }|x_1|<\cdots<|x_N|).
\eeq
Consequently ${\cal P}_\beta(x_1,\ldots,x_N)$ tends to 
a product of very narrow [of variance $\sigma^2=1/(4a\beta)\ll 1$]
Gaussian distributions 
obeyed by $x_i$ whose center is at $[\beta(i-1)+1]/2$.
This ``crystallization'' of eigenvalues invalidates naively expected
mutual independence of distributions of eigenvalues,
and drives the spectrum
toward an exotic statistics 
different from Poissonian \cite{BBP}.
This phenomenon can be rephrased in the context of
using the WKB formula (\ref{WKB}) to derive the kernel (\ref{KMCIN}).
Although Eq.\ (\ref{WKB}) remains valid even in the case $a\rightarrow\infty$,
use of the mean-field spectral density $\overline{\rho}(\l)=1/(2a \l)$ 
in place of the exact ${\rho}(\l)$
is not justifiable, because the crystallization of eigenvalues leads to 
a rapidly oscillating ${\rho}(\l)$.
In the case of the $q$-Hermite ensemble (\ref{Vq}),
the potential itself has a oscillation of the same type,
leading again to crystallization \cite{BBP}.
Therefore, the RME with log-squared potentials,
despite 
the fact that it is constructed from the macroscopic spectra of the AH, 
should be considered as a good model
of the latter only for small values of $a$ where
the level repulsion property of the Gaussian ensembles
is deformed slightly but not to the extent that
the crystallization of eigenvalues becomes prominent.
In the case of the $q$-Hermite ensembles,
we can estimate 
this scale to be characterized by the value of $a$ where
nontranslational invariance of the exact kernel
becomes manifest, i.e., $\e^{-\pi^2/a} \simeq 1$.
We assume this estimate to be valid generically 
for RME's with log-squared potentials (\ref{Vlog2}).
In view of this, for our purpose of reproducing the LSDF of the AH,
we shall concentrate on the approximated universal kernel (\ref{KMCIN}).
This will be justified a {\em posteriori}, after
confirming that the best-fit value of
the parameter $a$ for the MIT point of the AH's are
such that $\e^{-\pi^2/a} \ll 1$.

\section{level spacing distributions of the deformed kernel}
In this section we analytically compute the 
LSDF $P_\beta(s)$ from the deformed kernel (\ref{KMCIN})
for all values of the Dyson index $\beta$.
Our result completes earlier attempts to compute $P_2(s)$ numerically
\cite{MCIN} or asymptotically \cite{MK}, and is consistent with those.

We notice that the kernel (\ref{KMCIN})
is equivalent to that of 
Dyson's circular unitary ensemble
at finite $N$ \cite{Meh},
\beq
K(x,y)=\frac{\sin (N/2) (x-y)}{N \sin (1/2) (x-y)} ,
\label{KCUE}
\eeq
by the following analytic continuation
\beq
N\rightarrow \frac{\pi i}{a},\ \ 
x\rightarrow \frac{2a}{i}x .
\label{sub}
\eeq
Tracy and Widom \cite{TW94} have proven that the diagonal resolvent kernel
of Eq.\ (\ref{KCUE}) is determined by
a second-order differential
equation that is reduced to a Painlev\'e VI equation
\cite{DIZ}.
We shall follow their method below.

The kernel (\ref{KMCIN}) is written as
\bea
&&K(x,y)=\frac{\phi(x)\psi(y)-\psi(x)\phi(y)}{\e^{2ax}-\e^{2ay}},
\label{Kpp}\\
&&\phi(x)=\sqrt{\frac{2a}{\pi}}\,\e^{ax}\sin \pi x,\ \ 
\psi(x)=\sqrt{\frac{2a}{\pi}}\,\e^{ax}\cos \pi x.\nonumber
\eea
These component functions satisfy 
\beq
\phi'=a \phi + \pi \psi,\ \ 
\psi'=-\pi \phi + a \psi.
\eeq 
We use the bra-ket notation $\phi(x)=\langle x | \phi \rangle$
and so forth \cite{BH}. 
Due to our choice of the component functions
to be real valued (unlike \cite{TW94}, Sec.\ V D), we have 
$\langle x | {O} | \phi \rangle = \langle \phi | {O} | x \rangle$
and a similar situation for $\psi$
with any self-adjoint operator ${O}$ and real $x$.
Then Eq.\ (\ref{Kpp}) is equivalent to
\beq
[\e^{2aX}, K]=|\phi\rangle \langle \psi |-|\psi\rangle \langle \phi|,
\label{eK}
\eeq
where $X$ and $K$ are 
the multiplication operator
of the independent variable and the integral operator with the kernel
$K(x, y)\theta(y-t_1) \theta(t_2-y) $, respectively. 
Below we will not explicitly write the dependence 
on the end points of the underlying interval $[t_1,t_2]$.
It follows from Eq.\ (\ref{eK}) that
\beq
\left[ \e^{2aX}, \frac{K}{1-K} \right]=
\frac{1}{1-K}
( |\phi\rangle \langle \psi| - |\psi\rangle \langle \phi| )
\frac{1}{1-K} ,
\label{eR}
\eeq
that is, 
\bea
(\e^{2ax}-\e^{2ay}) R(x,y)&=& Q(x)P(y)-P(x)Q(y),
\label{Rxy}\\
Q(x)&\equiv&\langle x | (1-K)^{-1}| \phi \rangle,\nonumber\\
P(x)&\equiv&\langle x | (1-K)^{-1}| \psi \rangle. \nonumber
\eea
At a coincident point $x=y$ we have
\beq
2a\,\e^{2ax} R(x,x)= Q'(x)P(x)-P'(x)Q(x).
\label{Rxx}
\eeq
Now, by using the identity
\beq
\frac{\pa K}{\pa {t_i}}=(-1)^{i} K |t_i \rangle \langle t_i |
\ \ (i=1,2),
\eeq
we obtain
\begin{mathletters}
\label{dPQdti}
\bea
&&\frac{\pa Q(x)}{\pa {t_i}}=(-1)^{i} R(x, t_i) Q(t_i),
\label{dQdti}\\
&&\frac{\pa P(x)}{\pa {t_i}}=(-1)^{i} R(x, t_i) P(t_i).
\label{dPdti}
\eea
\end{mathletters}
On the other hand, by using the identity 
($D$ is the derivation operator)
\beq
[D,K]=K( |t_1 \rangle \langle t_1 |- |t_2 \rangle \langle t_2 |) ,
\eeq
which follows from the translational invariance of the kernel
$(\pa_x +\pa_y) K(x-y)=0$,
we also have 
\begin{mathletters}
\label{dPQdx}
\bea
\frac{\pa Q(x)}{\pa x}
&\!=\!&\langle x| D (1-K)^{-1} |\phi \rangle \nonumber\\
&\!=\!&\langle x| (1-K)^{-1} |\phi' \rangle \nonumber\\
&&+\langle x| (1-K)^{-1}[D,K](I-K)^{-1} |\phi \rangle \nonumber\\
&\!=\!&a Q(x) + \pi P(x) + R(x,t_1)Q(t_1)- R(x,t_2)Q(t_2),
\!\!\!\!\!\!\nonumber\\
&&\label{dQdx} \\
\frac{\pa P(x)}{\pa x}
&\!=\!&-\pi Q(x) + a P(x) + R(x,t_1)P(t_1)- R(x,t_2)P(t_2).
\!\!\!\!\!\!\nonumber\\
&&\label{dPdx}
\eea
\end{mathletters}
Now we set $t_1=-t$, $t_2=t$, $x, y= -t$ or $t$, and introduce notations
$\q=Q(-t), q=Q(t), \p=P(-t), p=P(t)$, and
$\R=R(-t,t)=R(t,-t), R=R(t,t)=R(-t,-t)$. 
The last two equalities follow from
the evenness of the kernel.
Then Eqs.\ (\ref{Rxy}) and (\ref{Rxx}) read, after using Eqs.\ (\ref{dPQdx}),
\begin{mathletters}
\label{pqpq}
\bea
&&  \p q-\q p= 2\R \sinh 2at ,\\
&&  \p^2+\q^2=\frac2\pi ( \R^2 \sinh 2at + R\, a\, \e^{-2at} ) , \\
&&   p^2+ q^2=\frac2\pi ( \R^2 \sinh 2at + R\, a\, \e^{ 2at} ) .
\eea
\end{mathletters}
The total $t$ derivatives of Eqs.\ (\ref{pqpq}) lead to
($\cdot = d/dt$)
\bea
&&  \p p+ \q q= \frac1\pi (\R \sinh 2at)^{\mbox{$\cdot$}},
\label{ppqq}\\
&& 
\dot{R}=2\R^2, \ \ \ddot{R}=4\R \dot{\R}.
\label{RR}
\eea
The left-hand sides of Eqs.\ (\ref{pqpq}) and (\ref{ppqq})
satisfy an additional constraint,
\beq
(\p p+\q q)^2+ (\p q-\q p)^2=
(\p^2 + \q^2) (p^2+q^2) .
\label{const}
\eeq
By eliminating $\p, p, \q, q$, $\tilde{R}$, and $\dot{\R}$ from 
Eqs.\ (\ref{pqpq})--(\ref{const}), 
we obtain for 
$R(s)$ ($s \equiv 2t$, $'=d/ds$)
\bea
&&\left[ a \cosh as\,R'(s)  \!+\! \frac{\sinh as}{2}R''(s) \right]^2
\!\! + \! [\pi \sinh as\,R'(s)]^2 =
\nonumber\\
&&R'(s)\left( [a R(s)]^2 \!+\!a\sinh 2as R(s)R'(s)
\!+\! [\sinh as\,R'(s)]^2 \right)\!.
\nonumber\\
&& \label{PVI} 
\eea
It is equivalent to Eq.\ (5.70) of Ref.\ \cite{TW94} after the
analytic continuation
(\ref{sub}), accompanied by a redefinition
$R(s)\rightarrow ({i}/{2a}) R(s)$.
This is slightly nontrivial because Ref.\ \cite{TW94} has used
$\p=p^*$ and $\q=q^*$, which follow from the
analytic properties of its component functions,
$\phi(-x)=\phi(x)^*$ and $\psi(-x)=\psi(x)^*$.
In the limit $a\rightarrow 0$.
it clearly reduces to the $\sigma$ form of a Painlev\'{e} V equation 
\bea
&&\left[R'(s)+\frac{s}{2}R''(s)\right]^2
+[\pi s R'(s)]^2= 
R'(s)[R(s) \!+\! s R'(s)]^2 ,\nonumber\\
&&\label{PV}
\eea
derived for the sine kernel (\ref{Ksin}) 
of the Gaussian ensembles \cite{JMMS}.
Note that for $\beta=1$, a replacement $a\rightarrow a/2$ is
necessary because of our convention (\ref{weight}).

Finally, the LSDF's $P_\beta(s)$
are expressed in terms of the diagonal resolvent $R(s)$ via 
Eqs.\ (\ref{E124}), (\ref{PE}), and the first of Eq.\ (\ref{RR}):
\bml
\label{P1P2P4}
\begin{eqnarray}
&&P_1(s)=\left[ \e^{ -\frac12 \int_0^s ds [R(s)+\sqrt{R'(s)}] } \right]''\!,\\
&&P_2(s)=\left[ \e^{ -\int_0^s ds R(s) } \right]'' \!,\\
&&P_4(s)=\left[
\e^{ -\frac12 \int_0^{2s} ds R(s) }
\cosh \left( \frac12 \int_0^{2s} ds \sqrt{R'(s)} \right) \right]'' \!\! .\!\!
\end{eqnarray}
\eml
The boundary condition follows from the expansion (\ref{Espert}) of $E_2(s)$
in terms of the correlation functions:
\bea
E_2(s)&=&1-\int_{-s/2}^{s/2} dx K(x,x) \nonumber\\
&&+\frac12 \int_{-s/2}^{s/2} dx_1\, dx_2 \det\limits_{i,j=1,2} 
K(x_i,x_j)-\cdots\nonumber\\
& =&1-s+O(s^4), \\
R(s)&=& -[\ln E_2(s)]'=1+s+\cdots
\label{bc}.
\eea
Our main result consists of
Eqs.\ (\ref{PVI}), (\ref{P1P2P4}), and (\ref{bc}).

For $s\ll 1/a$,
we can Taylor-expand hyperbolic functions,
to obtain a perturbative solution to Eq.\ (\ref{PVI}),
\widetext
\beq
R(s)=
1 + s + s^2
+ \left(1-\frac{\pi^2\!+\!a^2}{9}\right) s^3  
+ \left(1-\frac{5(\pi^2\!+\!a^2)}{36}\right) s^4  
+ \left(1-\frac{(\pi^2+a^2)(75 -4\pi^2-6a^2)}{450}\right) s^5
\!+\! \cdots,
\eeq
\vspace{-7mm}
\begin{mathletters}
\label{P124}
\bea
P_1(s)&=&
{\frac{ 4\,{{\pi }^2}+  {a^2} }{24}}s - 
  {\frac{\left(  4\,{{\pi }^2 + {a^2}} \right) \,
\left(  12\,{{\pi }^2 +7\,{a^2} } \right) }{2880}} s^3 + 
  {\frac{\left( {{\pi }^2}+ {a^2}\right) 
\,\left(  4\,{{\pi }^2+{a^2}} \right) }{1080}} s^4\nonumber\\
&&+ 
  {\frac{\left(  4\,{{\pi }^2}+{a^2} \right) \,
      \left( 
48\,{{\pi }^4}+ 72\,{{\pi }^2} \,{a^2}+ 31\,{a^4} 
 \right) \,{s^5}}{322560}} - 
  {\frac{\left( {{\pi }^2}+{a^2} \right) 
\,\left(  4\,{{\pi }^2}+ {a^2}\right) \,
      \left(12\,{{\pi }^2} +13\,{a^2} \right) \,{s^6}}{226800}}
+\cdots, \\
P_2(s)&=&
\frac{\pi^2  + {a^2}}{3}s^2
-\frac{(\pi^2 + {a^2})(2\pi^2 + 3{a^2})}{45} s^4
+\frac{(\pi^2 + {a^2})(\pi^2 + 2{a^2})(3\pi^2 + 5{a^2})}{945} s^6
+\cdots, \\
P_4(s)&=& 
{\frac{16\,\left(
 {{\pi }^2}+{a^2} \right) \,\left( {{\pi }^2}+ 4\,{a^2}\right) \,{s^4}}{135}}
 -  {\frac{128\,\left( {{\pi }^2}+ {a^2}\right) \,
\left(  {{\pi }^2}+  4\,{a^2}\right) \,
      \left(3\,{{\pi }^2} + 13\,{a^2} \right) \,{s^6}}{14175}}
+\cdots.
\eea
\end{mathletters}
\Rrule
\narrowtext
\noindent
The above perturbative expansions are 
correct for the kernel (\ref{Ktheta}) of the $q$-Hermite ensemble  
in any polynomial orders of $a$, while they lack
nonperturbative terms of order $\e^{-\pi^2/a}$,
which depend on the reference point.
On the other hand, for $s\gg 1/a$, 
it can be proven from Eq.\ (\ref{PVI})
that $R(s)$ approachs a constant.
Then Eqs.\ (\ref{P1P2P4}) imply
\bml
\label{P124poisson}
\bea
&&\ln P_{1}(s)\sim -\frac12 {R_{\frac{a}{2}}(\infty)} s,\\
&&\ln P_{2,4}(s)\sim -R_a(\infty) s,
\eea
\eml
for $s\rightarrow \infty$.
In Fig.\ 1 we exhibit the decay rate
$\kappa(a)\equiv R_a(s=\infty)$ for $0<a<4$ 
computed numerically from Eq.\ (\ref{PVI}).
At present we could not find an analytic form of
$\kappa(a)$.
For small $a(<0.5)$, it is well approximated by
$1/\kappa(a)\approx 0.202 a$, which agrees extremely well
with the value $2/\pi^2=0.2028$
expected from Eq.\ (\ref{chi}) and the analytic formula \cite{AZKS}
that holds in generality,
\beq
\chi=\frac{1}{2\kappa}.
\eeq
Eqs.\ (\ref{P124}) and (\ref{P124poisson}) tells that 
our LSDF's are indeed hybrids of 
the Wigner-Dyson-like 
[$P_\beta(s) \sim s^\beta$ for $s$ small]
and the Poisson-like distributions
[$P_\beta(s) \sim \e^{-\kappa s}$ for $s$ large].
In Figs.\ 2--4 we exhibit plots of the LSDF's
$P_\beta(s)$ for $\beta=1,2,4$
and for various $a$ such that $\e^{-{\pi^2}/{a}}\ll 1$, 
obtained by numerically solving Eq.\ (\ref{PVI}).

\section{Anderson Hamiltonians at MIT}
In this section we make comparison between
the LSDF's and the level number variance
in the exact diagonalization of the AH's
and our analytic results from multifractal RME's.

As numerical data to compare with, we adopt 
Ref.\ \cite{ZK97} for the AH [Eq.\ (\ref{AH1})] ($\beta=1$), 
Ref.\ \cite{BSZK} for the AH under a magnetic field (\ref{AH2})
with $\alpha=1/5$ ($\beta=2$), and
Ref.\ \cite{KSOS} for the AH with spin-orbit coupling (\ref{AH4})
with $\theta=\pi/6$ ($\beta=4$), at their MIT points.
We choose the best fit values of $a$ 
from the exponential decay rates $\kappa$ of the numerical data
using Fig.\ 1.
Based on the numerical results
$\kappa=1.9, 1.8, 1.7$ for $\beta=1, 2, 4$, respectively,
we estimate the parameter $a$ in the potentials of RME's to be
$a=2.95, 3.55, 3.90$.
Preference could alternately be put on
best matchings in the smaller values of $s$ $(\alt 2)$,
which would lead to $a=3.2$ for $\beta=1$
[although the difference in $P_1(s)$ 
between $a=2.95$ and $a=3.2$ is tiny].
In Figs.\ 5--7 we exhibit linear and logarithmic plots of LSDF's
of the RME's and the AH's.
The numerical data fit excellently with our analytic result 
from the kernel (\ref{KMCIN}),
for a large energy range $0\leq s \alt 6$ 
where the LSDF's vary by four to five orders of magnitude.
The use of this approximated kernel
is justifiable because $\e^{-\pi^2/a} \ll 1$ holds
for these values of $a$.
Small systematic deviations can be attributed
to the errors involved in determining the values of $a$ 
from the decay rates of numerical LSDF's, and possibly
to an essential difference of order $O(\e^{-\pi^2/a})$ between 
the RME's and the AH's.

Furthermore, 
we exhibit in Fig.\ 8 the number variance $\Sigma^2(S)$ 
of the orthogonal AH obtained in Ref.\ \cite{EK},
together with the RME result (\ref{Sigma2})
with $(\beta=1)$
\beq
Y_2(x)=K(x)^2+K'(x)\int_x^\infty K(y)dy
\eeq
at $a=3.2$.
We can confirm that not only the asymptotic slopes $\chi$
of $\Sigma^2(S)$
(first pointed out by Canali \cite{Can}, who computed $\chi$ 
by the Monte Carlo simulation of RME's),
but their full functional forms are in a good agreement
for $L \alt 10$.
To recapitulate, we have 
the following three distinct functional observables
(consisting of the correlation functions in the second column)
that agree well between the critical AH's and the deformed RME's:
\begin{center}
\begin{tabular}{lll}
Quantity & & ~~Correlation function\\[0.25ex]
\hline
Potential & $V(\lambda)$ & ~~1-level\\
Number variance & $\Sigma^2(S)$ & ~~unfolded 2-level\\
Level spacing & $P(s)$ & ~~unfolded $n$-level ($n\geq2$)\\
\end{tabular}
\end{center}
\noindent
In addition, both the critical AH's \cite{SG} and the deformed RME's \cite{KM}
are shown to have multifractal eigenstates, 
although the sequences of the multifractal dimensions 
are yet to be compared.
Agreements in the unfolded quantities
should not be considered a tautological
consequence of the first line;
one should recall that
an identical semicircle spectrum 
could as well be obtained either
from invariant 
RME's with Gaussian potentials (obeying the Wigner-Dyson statistics), 
or from diagonal random matrices whose entries are 
independently derived
from the semicircle distribution (obeying the Poisson statistics),
or from any intermediate ensembles.
From these grounds, we conclude that
the interaction between {\it unfolded}
energy levels of the 3D AH 
on equilateral ($L_x=L_y=L_z$) toroidal lattices
at the MIT is very well described by
the form $|\e^{2ax}-\e^{2ax'}|^\beta$,
which is common to 
the RME's with log-squared potentials (\ref{Vlog2}),
in contrast to the standard form $|x-x'|^\beta$ of the Gaussian RME's
and the AH's in the metallic regime.
We surmise that
the dimensionality and the fractal dimensionality
enter the critical spectral statistics
primarily through a single parameter $a$,
as long as the multifractality of the critical wave functions is
not too strong.
Further work is needed to explain this form of the level repulsion
from the multifractality (\ref{multi}) of the wave functions, and 
its origin from microscopic models.

Finally, remarks related to novel numerical results 
on critical AH's are in order.
Recently it was observed that the critical LSDF 
of the 3D AH is sensitive to the geometry of the lattice, i.e.,
the topology (boundary condition) \cite{BMP,SP} 
and the aspect ratio \cite{PS},
due to the coherence of the critical wave functions 
maintained over the whole lattice.
Since RME's treat randomness on all cites and bonds on equal footing,
it is likely that our RME's describe best the critical AH's 
on maximally symmetric lattices, i.e., equilateral tori,
but not those on less symmetric lattices, such as
unequilateral toroidal lattices
or lattices with boundaries.
Besides, since the validity of our expressions for the LSDF's 
is limited to the case of weak multifractality (relatively small $a$),
it will not properly describe the critical orthogonal AH in 
four dimensions \cite{ZK98},
where the level compressibility was observed to be 
larger than the value in three dimensions 
($\chi \approx 0.27$ \cite{ZK95}) and
close to its upper bound $\chi=0.5$.

\acknowledgments
I thank I.K. Zharekeshev, L. Schweitzer, T. Kawarabayashi,
S.N. Evangelou, E. Hofstetter, and C.M. Canali
for kindly providing me with their numerical data, and
H. Widom, J.T. Chalker, A. Zee, and E. Kanzieper 
for discussions and comments.
This work was supported in part by
JSPS Research Fellowships for Young Scientists,
by Grant-in-Aid No.\ 411044 
from the Ministry of Education, Science and Culture, Japan,
by the Nishina Memorial Foundation,
and by NSF Grant No.\ PHY94-07194,
and these supports are gratefully acknowledged.

\appendix
\section{Fredholm determinant in random matrix theory}
In the Appendix we collect known results on random matrix theories 
that are relevant to
our purpose of evaluating the LSDF of
three symmetry classes of RME's.
We follow Mehta's classical book \cite{Meh}
and the works by Tracy and Widom \cite{TW94,TW96,Wid}.
Readers are referred to them for detailed proofs. 
Subsequently we shall concentrate on the case where the spectral correlation
is translationally invariant after unfolding.

The joint probability densities of eigenvalues of $N\times N$ random
real symmetric ($\beta=1$), 
complex Hermitian ($\beta=2$), and
quaternion self-dual ($\beta=4$) matrices are given by
\bea
&&{\cal P}_\beta(x_1,\cdots,x_N)
={\rm const}\times
\prod_{i=1}^N w_\beta(x_i)\!\!
\prod_{1\leq i < j \leq N} \!\! \left| x_i-x_j \right|^\beta ,
\nonumber\\
&& \ w_{1}(x)=\e^{-V(x)/2},
\ \ 
w_{2,4}(x)=\e^{-V(x)}.
\label{weight}
\eea
We use the above convention between the weight functions
and the potentials
so as to simplify the relations between kernels [Eq.\ (\ref{S1S4}) below].
We introduce
the ``wave functions'' $\{ \psi_i(x) \}_{i=0,1,\ldots}$ by
orthonormalizing the sequence
$\{ x^i\, \e^{-V(x)/2} \}$, and
the projection operator $K$
to the subspace spanned by the first $N$ wave functions.
As an integration operator acting on the Hilbert space
spanned by 
the wave functions, $K$ is associated with the kernel
(we shall use the same letter for an operator 
and the kernel associated with it),
\beq
K(x,y)=\sum_{i=0}^{N-1} \psi_i(x) \psi_i(y).
\label{A3}
\eeq
Then the joint probability densities are expressed in terms of
determinants of the kernels \cite{Meh}:
\beq
{\cal P}_\beta(x_1,\ldots,x_N)=
\det\limits_{1\leq i,j \leq N}
K_\beta(x_i,x_j) , 
\label{P=det}
\eeq
\bml
\bea
&& K_2(x,y)=K(x,y),\\
&&K_1(x,y)=
\left(
\begin{array}{ll}
S_1(x,y) & S_1 D (x,y)\\
\epsilon S_1(x,y)-\epsilon(x, y) & S_1(y,x)
\end{array}
\right), 
\label{K1}\\
&&K_4(x,y)=
\left(
\begin{array}{ll}
S_4 (2x,2y)& S_4 D (2x,2y)\\
\epsilon S_4 (2x,2y)& S_4 (2y,2x)
\end{array}
\right) .
\label{K4}
\eea
\eml
Here $\det$ is to be interpreted as a quaternion determinant
in the case of $\beta=1$ and 4 \cite{Meh}. 
$D$ stands for the differentiation operator,
and $\epsilon$, $S_1$ and $S_4$ 
stand for integration operators
with kernels \cite{Wid}:
\bml
\label{S1S4}
\bea
\epsilon(x,y)&=&\frac12 {\rm sgn}(x-y), \nonumber\\
S_1 (x,y) &=& [1-(1-K)\epsilon K D]^{-1}K (y,x),
\label{S1}\\
S_4 (x,y) &=& [1-(1-K)D K \epsilon ]^{-1}K (x,y).
\label{S4}
\eea
\eml
A composite operator such as $\epsilon S_1$ 
is defined to have a convoluted kernel,
$\epsilon S_1(x,y)=\int_{-\infty}^\infty dz\, \epsilon(x,z) S_1(z,y)$
and so forth, and $[\cdots]^{-1}$ stands for an inverse operator.

The probability $E_\beta[J]$ of finding no eigenvalues in 
a set of intervals $J$
is defined as
\beq
E_\beta[J]=
\int_{x_i \not\in J}
 dx_1\cdots dx_N
{\cal P}_\beta(x_1,\ldots,x_N). 
\label{Es}
\eeq
By virtue of Eq.\ (\ref{P=det})
and the identity 
$\sum\limits_{\{n_i\}}\det\limits_{i,j} M_{n_i n_j}\propto 
\det\limits_{n,m} M_{n m}$, 
it is expressed in terms of the Fredholm determinant of 
the scalar or matrix kernel \cite{TW96}:
\bml
\label{Fredholm}
\bea
E_2[J]&=&\det( 1-\left. K_2 \right|_J),
\label{Fredholm2}\\
E_{1,4}[J]&=&\sqrt{\det( 1-\left. K_{1,4} \right|_J)},
\label{Fredholm14}\eea
\eml
where $|_J$ represents restriction of the kernel to the interval $J$.

In the following we assume that the scalar (unitary) kernel $K$ is
translationally invariant and symmetric
(the latter property is respected for invariant RME's by construction (\ref{A3}), 
but it is violated for RME's with partly 
deterministic matrix elements \cite{BH}),
\beq
K(x,y)=K(y,x)=K(x-y).
\label{trinvsym}
\eeq
If $K$ has these two properties, 
Eqs.\ (\ref{S1S4}) immediately reduce to 
\beq
S_1(x,y)=S_4(x,y)=K(x-y),
\eeq
because of the relations $DK=KD$ and $\epsilon D=1$, 
and the orthogonality of the two projection operators, $(1-K)K=0$.
We assume that $\lim_{N\rightarrow \infty}$ and the algebraic
manipulation that lead to the relation (\ref{S1S4})
can be interchanged.
In the following we let the notation $K$ represent its restriction to $J$, 
formerly denoted as $K|_J$.
If $J$ consists of a single interval $[-t,t]$,
we can simplify $E_\beta[J]$ significantly \cite{TW94}.
To do so, we introduce the resolvent operator $R=K(1-K)^{-1}$, and 
denote the associated kernel in the form of a matrix element,
\beq
R(x,y)\equiv \left\langle x \left| \frac{K}{1-K} \right| y \right\rangle.
\eeq
Due to the property (\ref{trinvsym}) assumed on the kernel,
it satisfies
\beq
R(x,y)=R(y,x)=R(-x,-y).
\eeq
Using this resolvent kernel, 
the Fredholm determinants (\ref{Fredholm}) are 
expressed as
\begin{mathletters}
\label{E124}
\bea
&&E_2[-t,t] =
\exp\left\{-2\int_0^t dt\, R(t,t) \right\},
\label{E124-2}\\
&&E_1[-t,t] =
\exp\left\{-\int_0^t dt [R(t,t) + R(-t,t)] \right\}, \label{E124-1}\\
&&E_4[-t,t]=
\frac12 
\left(
\exp\left\{-\int_0^{2t} dt [R(t,t) + R(-t,t)]\right\}\right.\nonumber\\
&& ~~~~~~~~~~\;
+\left. \exp\left\{-\int_0^{2t} dt [R(t,t) - R(-t,t)]\right\} \right).
\label{E124-4}
\eea
\end{mathletters}
Equation (\ref{E124-2}) can be proven by taking the logarithmic 
derivative of Eq.\ (\ref{Fredholm2}):
\bea
\frac{d}{dt} \ln E_2[-t,t]&=&
\tr \left( \frac{1}{1-K}\frac{dK}{dt} \right) \nonumber\\
&=&-\tr \left( \frac{1}{1-K} K 
(|-t\rangle\langle -t|+|t\rangle\langle t|) \right) \nonumber\\
&=&-2 \left\langle t \left|\frac{K}{1-K}\right| t \right\rangle.
\eea
Eqs.\ (A12b,c)
can be proven analogously from Eqs.\ (\ref{K1}), (\ref{K4}), and 
(\ref{Fredholm14}).
These relations are equivalent to Eqs.\ (6.5.19) and (10.7.5)
of Ref.\ \cite{Meh}
because $\exp\{-\int_0^t dt [R(t,t) \pm R(-t,t)] \}$ is a 
Fredholm determinant of the kernel $K(x,y) \pm K(-x,y)$
(although Ref.\ \cite{Meh} concerns primarily the sine kernel,
the proofs of these equations are equally valid
for any translationally invariant and symmetric kernel).\\
\indent
Now we set $2t=s$ and denote $E_\beta(s)=E_\beta[-s/2,s/2]$. 
The probability $P_\beta(s)$ for 
a pair of consecutive eigenvalues to have a spacing $s$ is 
clearly equal to the probability of finding 
an eigenvalue in an infinitesimal interval $[-s/2-\epsilon, s/2]$,
another in $[s/2, s/2+\epsilon']$ and none in between $[-s/2, s/2]$,
divided by $\epsilon\epsilon'$. Thus we have
\beq
P_\beta(s)=E_\beta''(s).
\label{PE}
\eeq

\newpage
\widetext
\begin{figure}
\epsfxsize=220pt
\begin{center}
\leavevmode
\epsfbox{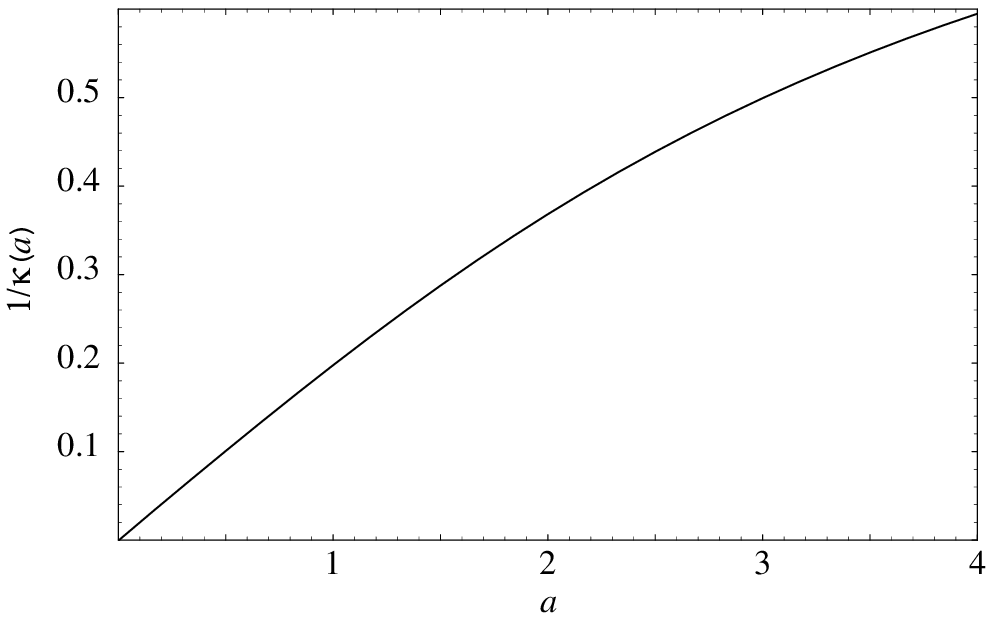}
\end{center}
\caption{
The decay rate $\kappa(a)=R_a(\infty)$ of the LSDF.}
\epsfxsize=230pt
\begin{center}
\leavevmode
\epsfbox{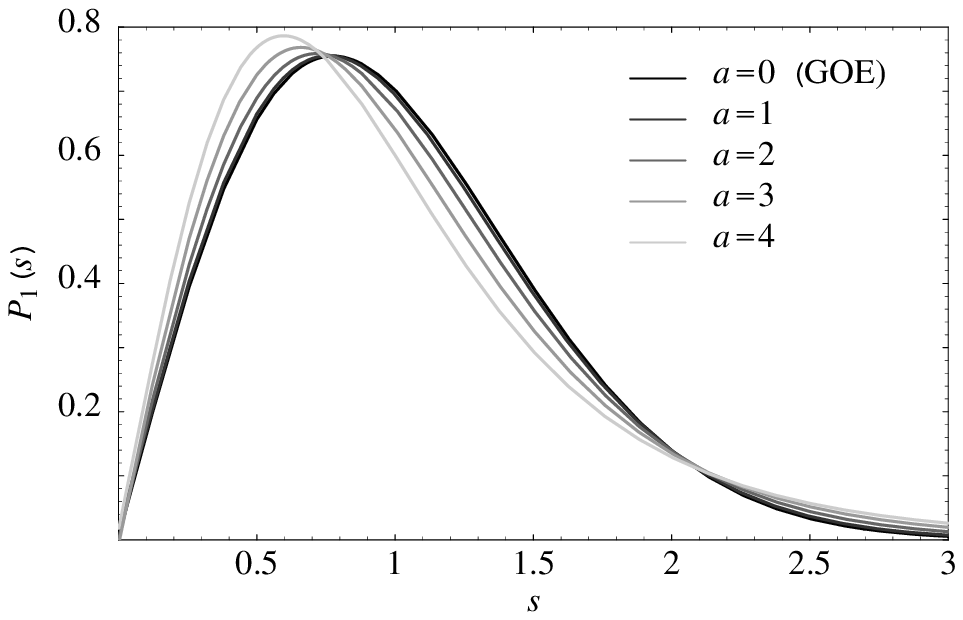}
\end{center}
\caption{
The LSDF $P_1(s)$ of the orthogonal ensemble
with the kernel (\ref{KMCIN}).}
\epsfxsize=230pt
\begin{center}
\leavevmode
\epsfbox{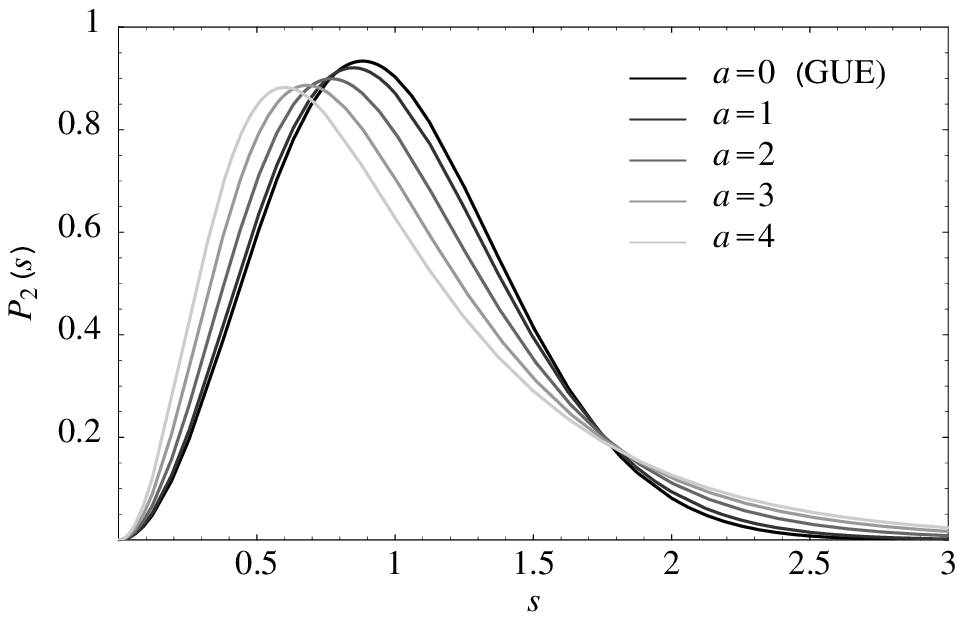}
\end{center}
\caption{
The LSDF $P_2(s)$ of the unitary ensemble
with the kernel (\ref{KMCIN}).}
\epsfxsize=230pt
\begin{center}
\leavevmode
\epsfbox{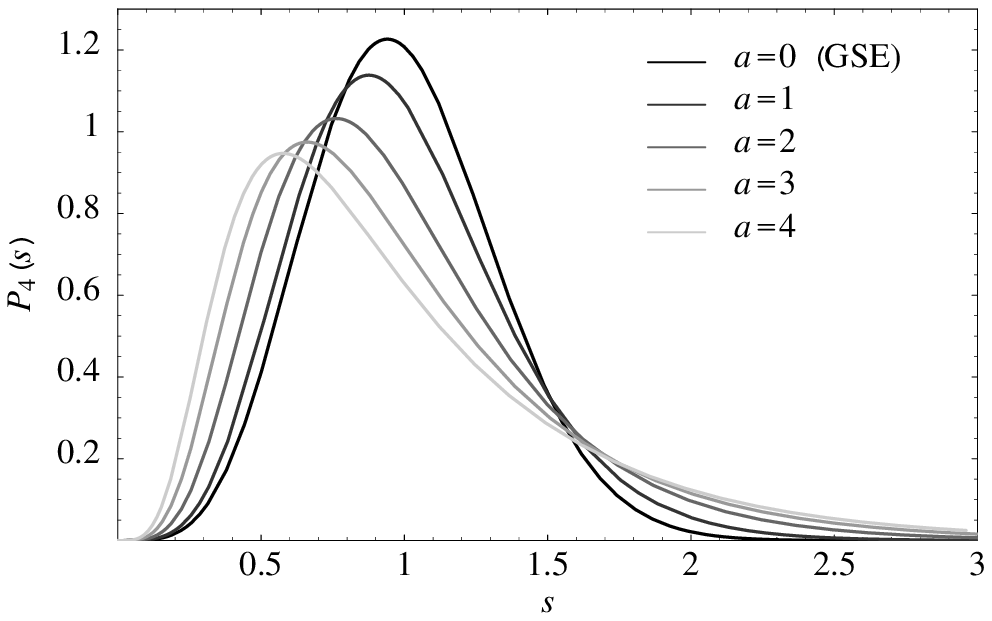}
\end{center}
\caption{
The LSDF $P_4(s)$ of the symplectic ensemble
with the kernel (\ref{KMCIN}).}
\end{figure}
\newpage
\begin{figure}
\begin{center}
\leavevmode
\epsfxsize=295pt
\epsfbox{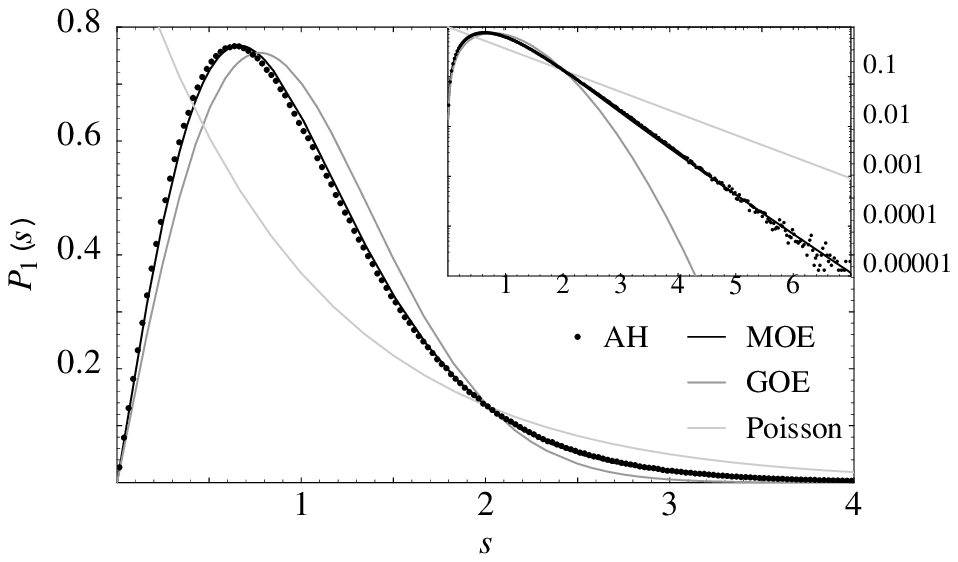}
\end{center}
\caption{
The LSDF's $P_1(s)$ of the multifractal orthogonal ensemble (MOE) at $a=2.95$
and of the 
Anderson Hamiltonian (\ref{AH1})
at the MIT point $W=16.4$, on a lattice 
of size $L^3=12^3$.
Numerical data are reprinted from Fig.\ 1 in Ref.\ \protect\cite{ZK97} 
courtesy of Zharekeshev.}
\begin{center}
\leavevmode
\epsfxsize=295pt
\epsfbox{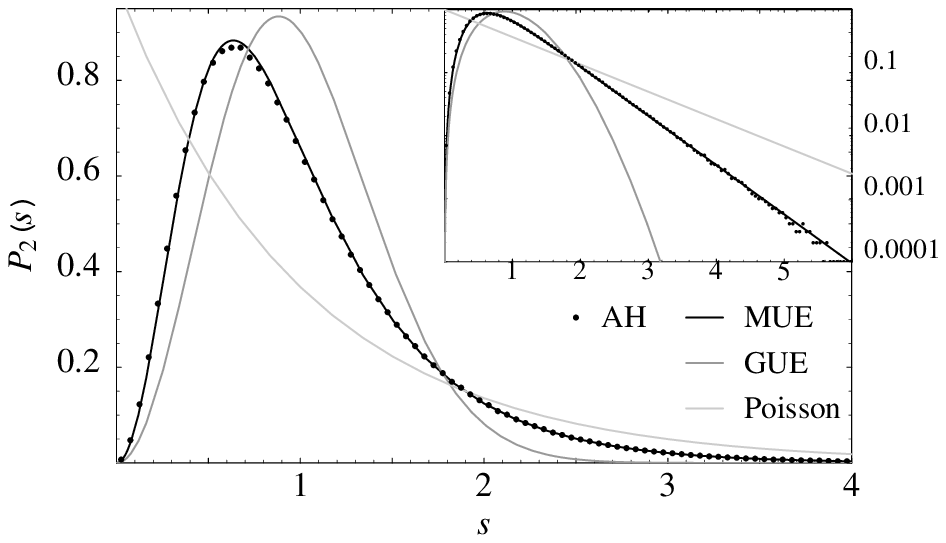}
\end{center}
\caption{
The LSDF's $P_2(s)$ of the multifractal unitary ensemble (MUE) at $a=3.55$
and of the AH (\ref{AH2})
under a magnetic field $\alpha=1/5$
at the MIT point $W=18.1$, on a lattice 
of size $L^3=5^3$.
Numerical data are reprinted from Fig.\ 1 in Ref.\ \protect\cite{BSZK} 
courtesy of Schweitzer.}
\begin{center}
\leavevmode
\epsfxsize=295pt
\epsfbox{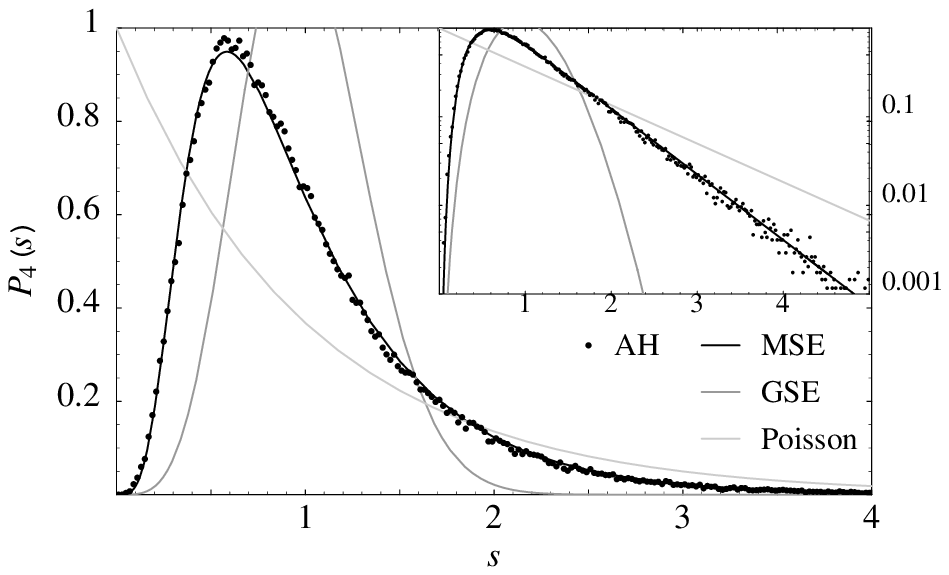}
\end{center}
\caption{
The LSDF's $P_4(s)$ of the multifractal symplectic ensemble (MSE) at $a=3.90$
and of the AH (\ref{AH4})
with spin-orbit coupling $\theta=\pi/6$
at the MIT point $W=19$,
on a lattice of size $L^3=12^3$.
Numerical data are reprinted from 
Figs.\ 2 and 3 in Ref.\ \protect\cite{KSOS} 
courtesy of Kawarabayashi.}
\end{figure}
\begin{figure}
\begin{center}
\leavevmode
\epsfxsize=260pt
\epsfbox{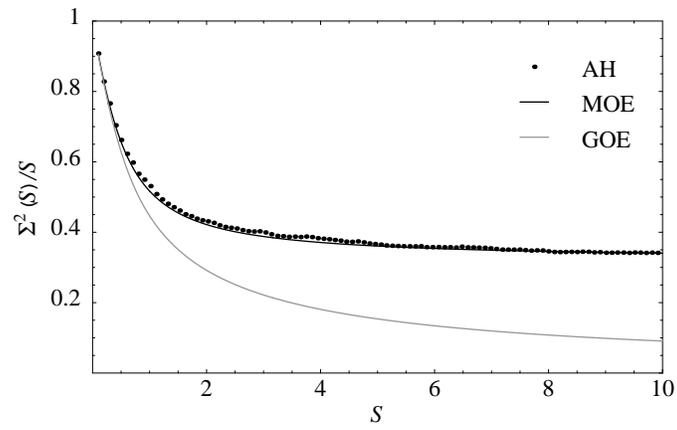}
\end{center}
\caption{
The level number variance $\Sigma^2(S)$ (divided by $S$) of 
the multifractal orthogonal ensemble at $a=3.2$
and of the AH (\ref{AH1}) at the MIT point $W=16.5$, 
on a lattice of size $L^3=10^3$. The Poisson distribution
corresponds to $\Sigma^2(S)/S=1$.
Numerical data are reprinted from Fig.\ 2 (b) in Ref.\ \protect\cite{EK}
courtesy of Evangelou.}
\end{figure}
\end{document}